\documentclass[aps,pra,twocolumn]{revtex4}
\usepackage[dvips]{epsfig}
\usepackage[usenames]{color}
\usepackage{pstricks}
\usepackage{amsmath}
\usepackage{amssymb}
\usepackage{subfigure}
\usepackage{afterpage}

\begin{document}

\title{Electric field suppression of ultracold confined
chemical rates}

\author{Goulven Qu{\'e}m{\'e}ner and John L. Bohn}
\affiliation{JILA, University of Colorado,
Boulder, C0 80309-0440, USA}

\date{\today}

\begin{abstract}
We consider ultracold collisions of polar molecules 
confined in a one dimensional optical lattice.
Using a quantum scattering formalism 
and a frame transformation method, we calculate elastic and
chemical quenching rate constants for fermionic molecules.
Taking $^{40}$K$^{87}$Rb molecules as a prototype,
we find that the rate of quenching collisions is {\it enhanced} at zero 
electric field as the confinement is increased, but that this rate is {\it suppressed}
when the electric field is turned on.
For molecules with 500~nK of collision energy,
for realistic molecular densities,
and for achievable experimental electric fields 
and trap confinements,
we predict lifetimes of KRb molecules of 1~s.
We find a ratio of elastic to quenching collision rates of about 100, 
which may be sufficient to achieve efficient experimental evaporative cooling
of polar KRb molecules.
\end{abstract}


\maketitle

\font\smallfont=cmr7

The recent achievement of ultracold 
polar molecules~\cite{Sage05,Ni08,Deiglmayr08}
has opened tremendous perspectives to the ultracold community.
The quantum states of the molecules 
can now be addressed experimentally~\cite{Ni08,Ospelkaus10-PRL}.
Beyond the idea of creating 
molecular Bose--Einstein condensates or degenerate Fermi gases of polar molecules,
these ``quantum-state controlled'' molecules have applications
in ultracold chemistry~\cite{Ospelkaus10-SCIENCE,Krems08-PCCP}, 
condensed matter and many-body physics~\cite{Micheli06-NATURE,Buchler07}, quantum information~\cite{Demille02,Yelin06}
or in precision measurement and test of fundamental laws~\cite{Carr09}.
These applications crucially depend on the collisional properties of the polar molecules
in the presence of an electric field.
To get a stable gas of molecules, elastic processes
should dominate over quenching (including inelastic and/or reactive) processes. 
This will be the case for certain molecules for which
chemical reactions are energetically forbidden
in their absolute ground state.
For others, as KRb molecules, chemical reactions
such as KRb + KRb $\to$ K$_2$ + Rb$_2$
are energetically allowed~\cite{Ospelkaus10-SCIENCE,Zuchowski10,Meyer10},
and evaporative cooling might be difficult to achieve.
Chemical rates
increase 
as the sixth power of the dipole moment induced by
an electric field, due to head-to-tail collisions of the polar molecules~\cite{Ni10-NATURE,Quemener10-QT}.  
On the other hand, these collisions may be
suppressed in a one dimensional (1D) 
optical lattice~\cite{Buchler07,Micheli07} that confines the molecules in planes perpendicular to the axis of their polarization, 
so that the dipoles mutually repel each other.  

In this Letter, we extend
the quantum formalism
used in Ref.~\cite{Ni10-NATURE}
to describe ultracold collisions
of polar molecules 
in an electric field
confined in a 1D optical lattice.
We apply the theory to predict
the elastic and chemical rate of confined KRb + KRb $\to$ K$_2$ + Rb$_2$ reactions
for which particular
attention is devoted~\cite{Ni08,Ospelkaus10-PRL,Ospelkaus10-SCIENCE,Ni10-NATURE}
and for which theoretical predictions are needed.
We show that the electric field suppression of confined chemical rates can help
to achieve efficient evaporative cooling of such molecules.
We assume that 
the wells of the 1D optical lattice 
are deep enough so that the molecules cannot tunnel
from site to site
and can be approximated by independent harmonic oscillator traps.
In the following, quantities are expressed in S.I. units, unless explicitly stated otherwise.
Atomic units (a.u.) are obtained by setting $\hbar = 4 \pi \varepsilon_0 = 1$.

We consider pairs of bosonic or fermionic polar molecules of mass $m$ and reduced mass
$\mu = m/2$ 
confined in a harmonic oscillator trapping
potential $V_\text{ho}= \mu \, \omega^2 \, z^2/2$, which allows the molecules
to collide in a two dimensional (2D) configuration space.  
The electric field is applied
along the $z$ direction, perpendicular to the planes of free motion
in the trap.  
The Hamiltonian in the relative coordinate ${\vec R}$
is given by
\begin{eqnarray}
 H  =  - \frac{\hbar^2} {2 \mu } \nabla^2 + V_\text{ho}
+ V_\text{vdW} + V_\text{dd} +  V_\text{abs}.
\label{Hamiltonian}
\end{eqnarray}
The long-range interactions are represented by the van der Waals potential
$V_\text{vdW} = - C_6 / R^6$
and  the dipole-dipole interaction
$V_{dd}= d^2 (1 - 3 \cos^2{\theta}) / (4 \pi \varepsilon_0 \, R^3)$, where
$d$ represents the effective dipole moment of the polar molecule
induced by the electric field~\cite{Quemener10-QT}.
Moreover, we represent chemical reactions via the non-Hermitian, 
absorbing potential $V_\text{abs}$
which has been chosen to
successfully describe three dimensional (3D) chemical rates 
of KRb molecules in an electric field, measured in Ref.~\cite{Ni10-NATURE}.

Scattering wave functions $\Psi$ are most naturally described at
large intermolecular spacing in cylindrical coordinates 
${\vec R}=(\rho, z, \varphi)$
in accordance with the symmetry of the confined trapping potential.
The interactions between molecules when they are closer together, however,
are better described in spherical coordinates ${\vec R} = (R, \theta, \varphi)$.
In our numerical calculations, we will therefore use the appropriate coordinate
system and then weld the wave functions together at a suitable matching 
distance.  Note that the azimuthal angle $\varphi$ is common to both
coordinate systems, and in fact the Hamiltonian is independent of $\varphi$.
Therefore, the quantum number $M$, representing the azimuthal projection
of the orbital angular momentum, is rigorously conserved.  Note also that in the limit 
of strong confinement, $\theta$ is restricted to values near 
$\theta \approx \pi/2$, in which case the dipolar interaction is repulsive.
This is the primary principle behind the electric suppression of collisions.

When the molecules are close together, we solve the coupled-channel
equations of motion in spherical coordinates using the diabatic-by-sector 
method~\cite{Pack87,Quemener06-PHD}.  
We divide the complete range of $R$ into sectors labeled by an index $p$.  In
each sector, we expand the $M$-dependent wave function of an initial channel $i$ as 
\begin{eqnarray}
\Psi^M_{i}(R,\theta) &=&  \frac{1}{R} \, \sum_{j} \chi^M_{j}(R_p;\theta) \, F^M_{j i}(R_p;R) .
\label{Psisph}
\end{eqnarray}
The adiabatic functions $\chi^M_j(R_p;\theta)$ are those that diagonalize the
$M-$dependent angular part ${\cal H}^M(R, \theta)$
of the Hamiltonian in Eq.~\eqref{Hamiltonian} 
at the fixed radius $R=R_p$, middle of the sector $p$.
The associated adiabatic energy is $\epsilon_j(R_p)$, which
converges to a harmonic oscillator
energy $\varepsilon_n$ of a state $n$ of the trap at large $R_p$.
The radial functions 
$F^M_{f i}(R_p;R)$, where $f=1,2,...$ represents an arbitrary final channel of the system,
are determined within each sector according to the
diabatic equations of motion
\begin{multline}
\left\{ - \frac{\hbar^2}{2 \mu} \frac{d^2}{d R^2} 
 - E \right\} 
\, F^M_{f  i}(R_p;R)  \\
+ \sum_{j} 
  {\cal U}^M_{f  j}(R_p;R) 
\  F^M_{j  i}(R_p;R) = 0
\label{eqcoup}
\end{multline}
where
\begin{multline}
{\cal U}^M_{f  j}(R_p;R) 
 =  \\
\int_0^{\pi} \, 
\chi^M_{f}(R_p;\theta) 
\, {\cal H}^M(R, \theta)
\, \chi^M_{j}(R_p;\theta)  \, \sin \theta \, d\theta . 
\label{Umatrix}
\end{multline}
$E$ is the total energy of the system.
To solve Eq.~\eqref{eqcoup} we employ the method of the propagation of the 
logarithmic derivative 
matrix $Z^M = (F^M)^{-1} (F^M)'$ of Johnson~\cite{Johnson73} 
up to a suitable matching radius $R_m$.

If $R_m$ is sufficiently large, then the only potential energy of any 
significance is the trap confinement potential $V_\text{ho}$.  At this point
the wave function is more conveniently described in cylindrical
coordinates.  
Therefore for an initial state $n_i$ of the trap, we expand the $M$-dependent wave function as
\begin{eqnarray}
\Psi^M_{n_i}(\rho,z) &=&  \frac{1}{\rho^{1/2}} \, \sum_{n_f} g_{n_f}(z) \, G^M_{n_f  n_i}(\rho) 
\label{Psicyl}
\end{eqnarray}
where $g_{n_f}(z)$ are normalized harmonic oscillator functions in $z$ 
corresponding to a state $n_f$ of the trap,
and represent the functions $ \rho^{1/2} \, \chi^M_f / R$ in the asymptotic limit.
The radial functions $G^M$ serve to define the 
reactance matrix $K^M$ via
$G^M_{n_f  n_i}(\rho) \propto \rho^{1/2} \, J_M(k_{n_f} \rho) \, \delta_{n_f n_i} + K^M_{n_f  n_i} \  \rho^{1/2} \, N_M(k_{n_f} \rho)$.
$J_M, N_M$ are Bessel functions. For closed channels, the modified Bessel functions has to be used instead.
$k_{n_f}=\sqrt{2 \, \mu \, (E - \varepsilon_{n_f})}/\hbar$
represents the wave-vector of the relative motion
of the state $n_f$.  
The $K^M$ matrix is found by a matching procedure
from the spherical wave function
that captures the short-range physics and the cylindrical wave function
that captures the asymptotic boundary conditions.
This is done by equating Eq.~\eqref{Psisph} to Eq.~\eqref{Psicyl}
at a constant radius $R=R_m$~\cite{Pack87,Quemener06-PHD},
and taking into account the one-to-one correspondence between the
short-range adiabatic states $i,f$ and their long-range counterparts
$n_i,n_f$.

The $K^M$ matrix in turn determines the scattering matrix 
$S^M = (I + iK^M)^{-1} (I - iK^M)$, 
where $I$ represents a diagonal unit matrix.
The cross sections for elastic and quenching collisions are
given by~\cite{Lapidus82,Adhikari86,Naidon06}
\begin{eqnarray}
\sigma^{el} &=&  \frac{\hbar}{\sqrt{2 \mu E_c}} \sum_{M} |1-S^M_{n_i n_i}|^2 \times \Delta   \\ 
\sigma^{qu} &=&  \frac{\hbar}{\sqrt{2 \mu E_c}} \sum_{M} \bigg( 1 - |S^M_{n_i n_i}|^2 \bigg) \times \Delta    
\label{crossrate}
\end{eqnarray}
and the rate coefficient is given by 
${\cal K}^{el,qu}  =  \sigma^{el,qu} \times  v$, where $v = \sqrt{2 \, E_c / \mu}$ is the
collision velocity.
The factor $\Delta$
represents symmetrization requirements for identical particles:
$\Delta=1,2$ according as the particles are distinguishable or 
indistinguishable.
To compare with experimental results, one should average the rate coefficients
over a Maxwell-Boltzmann distribution of the velocity $v$,
but we do not perform this average here.
Finally, due to exchange symmetry of indistinguishable molecules,
the quantum numbers in the asymptotic representation Eq.~\eqref{Psicyl}
satisfy the relation 
\begin{eqnarray}
(-1)^{n+M} = \gamma
\label{sym}
\end{eqnarray}
with $\gamma=+1$ for bosonic molecules
and $\gamma=-1$ for fermionic molecules.

\begin{table}[t]
\begin{center}
\begin{tabular}{c c c}
\hline
interaction & $a_\text{vdW,dd,ho}$ (a$_0$) & $E_\text{vdw,dd,ho}$ ($\mu$K)  \\ [0.5ex]
\hline
$C_6 = 21000$~a$_0$ & 264 & 19.6 \\ [0.5ex]
\hline
$d=0.1~\text{D}$  & 179 & 85.2 \\
$d=0.3~\text{D}$  & 1611 & 1.06  \\
$d=0.566~\text{D}$  & 5734 & 0.083 \\ [0.5ex]
\hline
$\nu=50~\text{kHz}$  & 1069 & 2.4 \\
$\nu=1000~\text{kHz}$  & 239 &  48 \\ [1ex]
\hline
\end{tabular}
\end{center}
\caption{Characteristic lengths and energies
of the different interactions
involved in the KRb + KRb collision.
1~D = 1~Debye = $3.336 \, 10^{-30}$~C m.
\label{TAB1}
}
\end{table}

For concreteness, we now consider collisions of fermionic 
$^{40}$K$^{87}$Rb molecules, prepared in indistinguishable internal states.  
We use the
actual mass and permanent dipole moment of these molecules, and set the coefficient
of their van der Waals interaction to
$C_6 = 21000 \, a.u.$~\cite{Ni10-NATURE} (1~a.u. = 1~$E_h \, a_0^6$
where $E_h$ is the Hartree energy and $a_0$ is the Bohr radius).
We consider only molecules initially confined to the ground state of
the harmonic oscillator, with $n_i=n_f=0$, and we take the collision
energy to be $E_c=500$~nK, relevant in the ongoing KRb experiment.
Under these circumstances, we find converged results at a matching
radius $R_m = 10^4$ $a_0$.  
As the pair of fermionic molecules is in the ground harmonic oscillator
state $n=0$, only odd values of $M$ are allowed in Eq.~\eqref{sym}.
This circumstance removes the occurrence of undesired head-to-tail 
collisions of the molecules in an electric field at ultralow energy.
At $E_c=500$~nK, we find that scattering is largely dominated by
the single partial wave $M=1$, similar to the 3D 
case~\cite{Ni10-NATURE,Quemener10-QT}.  
Despite the fact that the molecules are in the ground state of the trap,
we need to employ one asymptotically open channel ($n=0$)
and three asymptotically closed channels ($n=2,4,6$)
to converge the results to 10~$\%$.

The characteristic lengths ($a$) and energies ($E$)
of the different interactions involved in the chemical process
are presented in Table~\ref{TAB1}.
They are given for the various interactions by
$a_\text{vdW} = (2 \, \mu \, C_6 / \hbar^2)^{1/4}$ 
and $E_\text{vdW} = \hbar^2/(2 \, \mu \, a^2_\text{vdW})$   
for the van der Waals interaction,
$a_\text{dd}(d) = \mu \, d^2 / \hbar^2$ and 
$E_\text{dd}(d) = \hbar^6/(\mu^3 \, d^4)$ 
for the dipole-dipole interaction,
and
$a_\text{ho}(\nu) = \sqrt{\hbar/(2 \pi \nu \, \mu)} = \sqrt{\hbar/(\omega \, \mu)}$ 
and 
$E_\text{ho}(\nu) = \hbar \, 2 \pi \nu = \hbar \omega$ 
for the harmonic oscillator trap.
This is useful for characterizing the different regimes
involved in the chemical process.
When $a_\text{vdW} \ \text{or} \ a_\text{dd} < a_\text{ho}$,
intermolecular forces 
take place where the confinement is small
and the collisions are effectively 3D.
When
$a_\text{ho} < a_\text{vdW} \ \text{or} \ a_\text{dd}$, the reverse is true
and the collisions are effectively 2D.
For fermions in $n=0$ and when an electric field is applied, the molecules
meet primarily side-by-side and repel each other.
For weakly polarized molecules, the 3D and 2D limits are
realized for trap frequencies of 50~kHz and 1000~kHz, 
respectively (see Table~\ref{TAB1}).
For stronger polarized molecules at $d > 0.3$~D,
the 2D limit is reached for both confinements as $a_\text{ho} < a_\text{dd}$.
We consider the two confinements in the following.
We note that the present theoretical formalism
can treat both 3D
and 2D limits in an electric field. 
This is in contrast with 
former theoretical studies~\cite{Petrov01,Li09}
which do not describe the electric field dependence
and the 2D limit where
$a_\text{ho} < a_\text{vdW} \ \text{or} \ a_\text{dd}$.

\begin{figure} [h]
\begin{center}
\includegraphics*[width=7.5cm,keepaspectratio=true,angle=0]{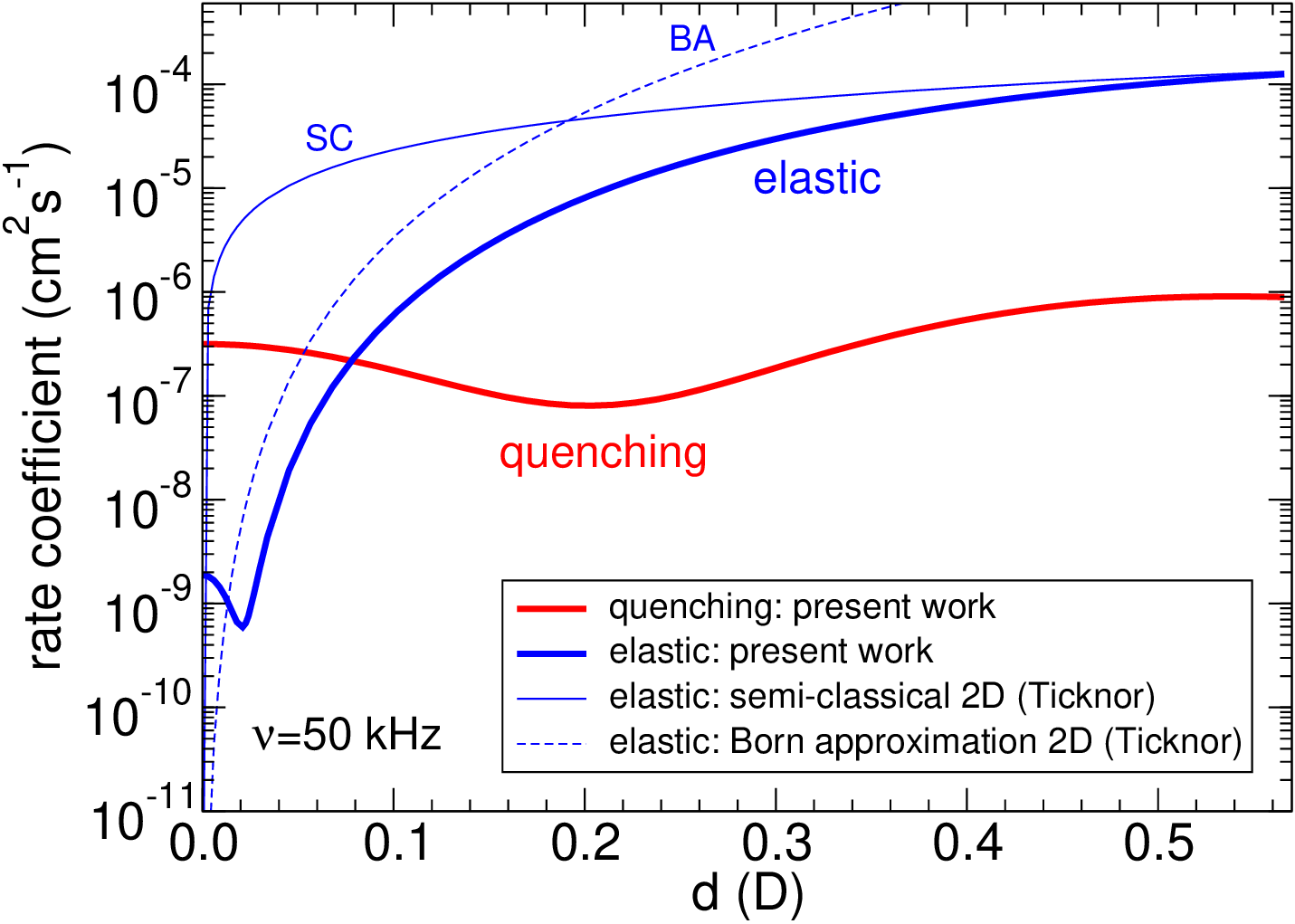} \\
\includegraphics*[width=7.5cm,keepaspectratio=true,angle=0]{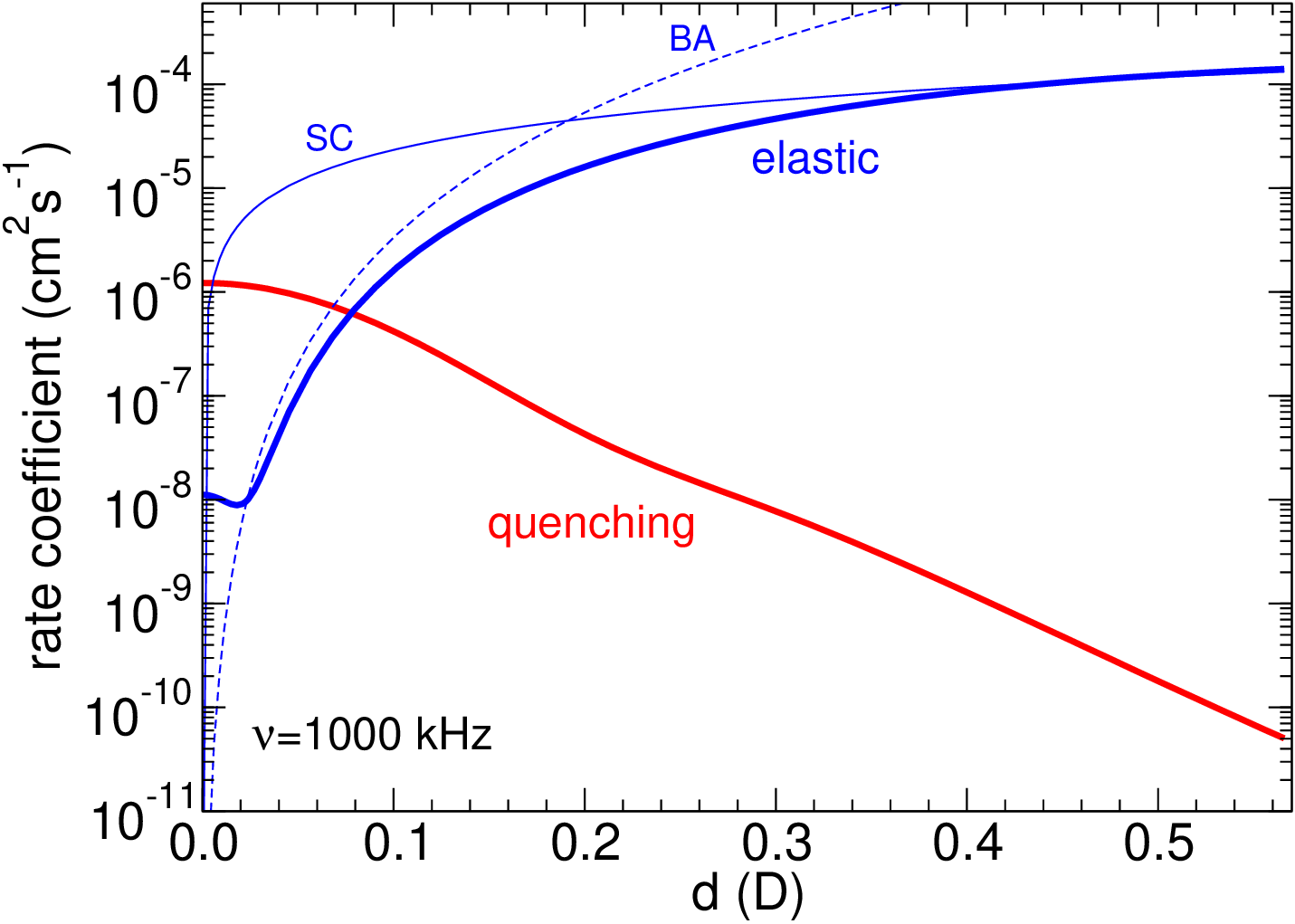}
\caption{(Color online) 
Collision of 
$^{40}$K$^{87}$Rb molecules in the ground state of a confining trap.
The elastic (blue) and quenching (red) rate coefficients 
are plotted as a function of the dipole moment $d$
for $\nu=50$~kHz (upper panel)
and $\nu=1000$~kHz (lower panel), for a fixed collision energy of $E_c=500$~nK.
A pure 2D semi-classical (SC) formula (thin blue line)
and a pure 2D Born approximation (BA) formula (dashed blue line)
has also been reported from Ticknor~\cite{Ticknor09}.
\label{EDIP-FIG}
}
\end{center}
\end{figure}

We present in Fig.~\ref{EDIP-FIG}
the elastic and quenching rate coefficient
as a function of the induced dipole moment $d$,
for a trap frequency of $\nu=50$~kHz (upper panel)
and $\nu=1000$~kHz (lower panel).
The elastic rates (blue lines) have the same trend 
for the two different confinements, increasing with the dipole moment.
For $d > 0.2$~D, the elastic rates converge
to a semi-classical formulation (thin blue lines)
of pure 2D dipolar scattering~\cite{Ticknor09},
which increases linearly with $d$.
For $d < 0.2$~D, the elastic rates are in better agreement 
with a pure 2D Born approximation (dashed blue lines)
that scales as $d^4$~\cite{Ticknor09}.
For $d \simeq 0$~D, the elastic rates take a finite value which is determined
by an unknown scattering phase shift
which depends on the short-range potential of KRb--KRb. 
We note that the overall elastic rate is in better agreement
with the pure 2D estimations
for the larger confinement $\nu=1000$~kHz than for the smaller confinement $\nu=50$~kHz, as one expects.

In contrast,
the quenching rates (red lines) have a different trend with dipole moment
for the two different confinement strengths.
For $\nu=50$~kHz (upper panel), the quenching rate first decreases and then increases again
as a function of $d$.
This behavior already contrasts with the quenching rate in 3D collisions,
which increases as $d^6$ in the absence of $z$-confinement~\cite{Ni10-NATURE,Quemener10-QT}.
The crucial difference comes from the fact that destructive head-to-tail collisions
are removed by the confined geometry for fermions in the ground state $n=0$ (see Eq.~\eqref{sym}).  
Under even greater confinement (lower panel)
the quenching rate continues to decrease
with increasing dipole moment~\cite{Quemener10-QT,Ticknor10},
illustrating the electric field suppression of confined chemical rates.

\begin{figure} [t]
\begin{center}
\includegraphics*[width=8cm,keepaspectratio=true,angle=0]{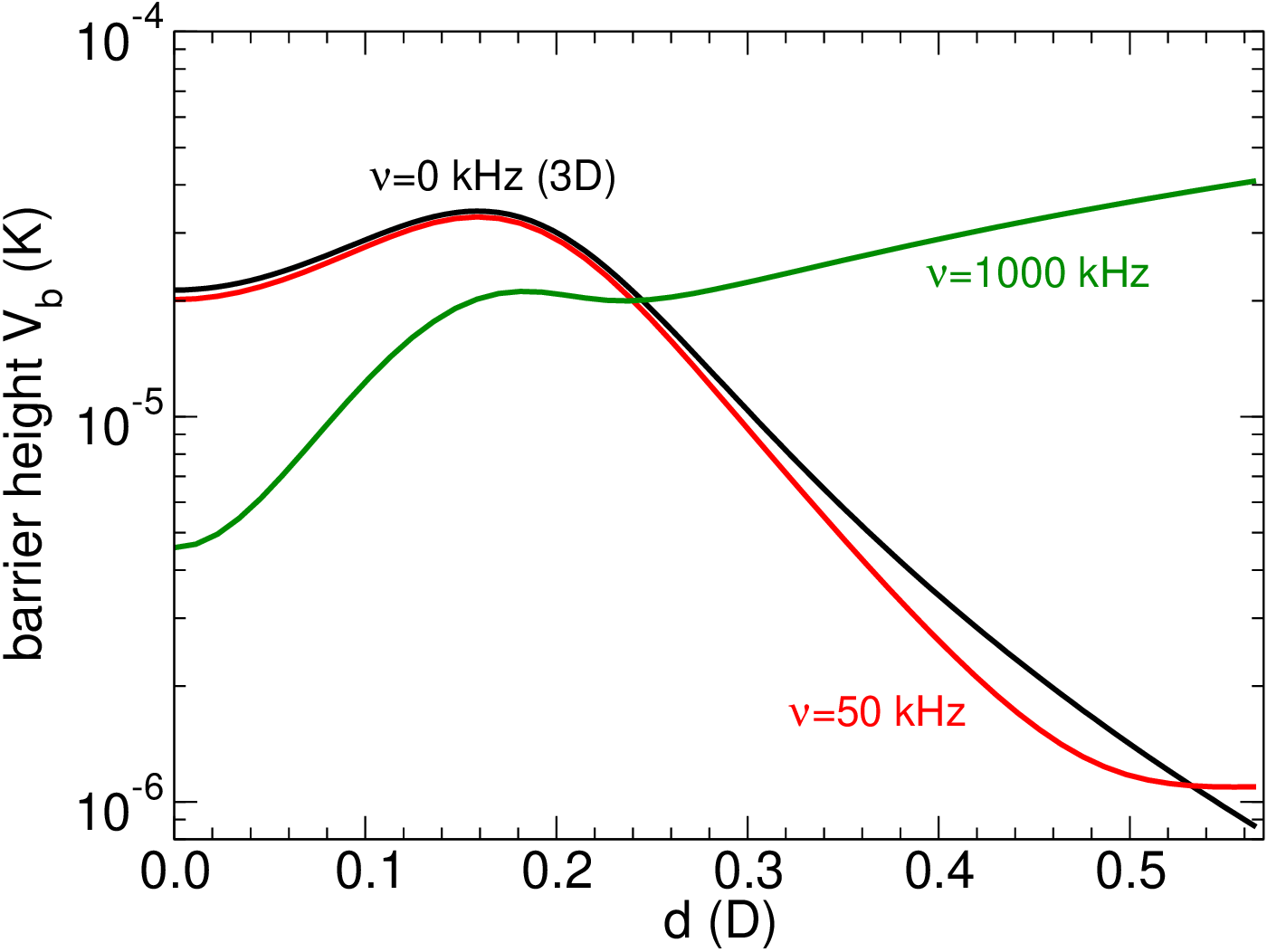}
\caption{(Color online)
Barrier heights $V_b$ as a function of the dipole moment $d$ for 
$\nu=0$~kHz (black),
$\nu=50$~kHz (red)
and $\nu=1000$~kHz (green).
\label{BARRIER-FIG}
}
\end{center}
\end{figure}

To better understand the qualitative trend of the quenching rates,
we present in Fig.~\ref{BARRIER-FIG}
the height $V_b$ of the effective potential energy barrier
corresponding to the
lowest adiabatic energy curve $\epsilon_{f=1}(R_p)$,
as a function of $d$
for $\nu=0,50,1000$~kHz.
We follow the qualitative arguments given in Ref.~\cite{Quemener10-QT}
that the behavior of the quenching rates  are suppressed by the
need for the molecules to tunnel through this barrier to the region of
chemical reactivity.
For $\nu=50$~kHz, the barrier increases for small dipole moments, 
since the dipole interaction is repulsive for $M=1$.  However,
anisotropy of the interaction couples different channels together.
Therefore, at higher dipole moments, the lowest adiabatic curve
is repelled more strongly from the others, ultimately lowering the barrier
height and increasing the quenching rate.
For the stronger confinement $\nu=1000$~kHz, the barrier height
continues to grow as the dipole moment increases, leading to
continued suppression of quenching, by four orders of magnitude
over the range of dipole moment shown.  
There are two ways to see this ongoing suppression of quenching.
First, since $a_\text{ho}(1000~\text{kHz}) < a_\text{vdW} \ \text{or} \ a_\text{dd} $,
the molecules exert strong dipolar forces on each other at long range
where they are still described by harmonic oscillator states.
Thus the interaction remains overwhelmingly repulsive and the
molecules do not get close enough to react.
Alternatively, we note that the trap confinement is tight enough 
so that the spacing 
between the harmonic trap levels is large, and  therefore the repulsion between 
adiabatic channels is small.  The incident adiabatic channel
is therefore less likely to be lowered due to the presence of other locally
open adiabatic channels at short range.
As a result, the barrier height continues to increase as $d$ increases,
driven solely by the increasing dipolar repulsion.

The calculations show another interesting trend as confinement is increased.
Namely, at zero dipole moment the quenching rate is higher
for the tighter confinement $\nu = 1000$~kHz.
This is consistent with the barrier height $V_b$ being smaller for the higher
trap frequency (Fig.~\ref{BARRIER-FIG}).
This difference arises from the fact that the collision energy is
measured relative to the asymptotic energy of the potential, which is the
zero point energy of the confining potential.  The tighter trap
has a higher zero-point energy (Tab.~\ref{TAB1}), hence the apparent barrier is lower.

Finally, Fig.~\ref{EDIP-FIG} 
shows that for an experimentally achievable frequency trap of $\nu=50$~kHz (upper panel),
we predict a loss rate of $\sim 10^{-7}$~cm$^{2}$~s$^{-1}$ per molecule
for the maximum dipole moment $d=0.2$~D achieved so far in the KRb
experiment.
For a realistic planar density of molecules $\sim 10^{7}$~cm$^{-2}$,
this corresponds to molecular lifetimes of $\sim 1$~s which is 100 times longer
than if the molecules were not confined.
More important is the number of elastic collisions
per chemical reaction, given by the ratio of elastic to 
quenching rates.
At $d=0.2$~D for $\nu=50$~kHz, we predict $\sim 100$ elastic collisions per chemical reaction.
This ratio may be sufficiently high to achieve efficient evaporative cooling
of KRb molecules in a 1D optical lattice.

We acknowledge the financial support
of NIST, the NSF, and an AFOSR MURI grant.
We thank the JILA experimentalists
D. Wang, M. H. G. de Miranda, B. Neyenhuis, A. Chotia,
K.-K. Ni, S. Ospelkaus, J. Ye, D. S. Jin,
and the JILA theorists J. P. D'Incao, C. H. Greene, A. M. Rey
for stimulating discussions.
We also thank P. S. Julienne, S. Ronen, G. Pupillo, A. Micheli, P. Zoller, C. Ticknor
for helpful discussions.

\end{document}